\documentclass{article}
\usepackage{mrs2005,epsfig}
\begin{document} 
\title{Luminosity Functions and Star Formation Rates at $z\sim6-10$:
 Galaxy Buildup in the Reionization Age}

\author{Rychard Bouwens and Garth Illingworth}
\affil{Department of Astronomy, University of California, Santa Cruz}

\begin{abstract} 
HST ACS and NICMOS data are now of sufficient depth and areal coverage
to place strong constraints on the formation and evolution of galaxies
during the first 1-2 Gyrs of the universe.  Of particular interest are
galaxies at $z\sim6$ since they represent the earliest epoch
accessible to current high-efficiency optical instrumentation.  To
this end, we have constructed a sample of 506 $z\sim6$ objects from
all the deepest wide-area HST data (UDF, UDF-Parallel, and GOODS
fields).  They have been used to construct an optimal determination of
the rest-frame continuum UV LF at $z\sim6$.  Our LF extends to over 3
magnitudes below $L^{*}$, fainter than has been done at $z\sim3$.
Over the interval $z\sim6$ to $z\sim3$, we find strong evidence for
evolution in the UV LF.  Though constraints on the faint-end slope
remain modest (and are consistent with no-evolution), the
characteristic luminosity appears to have approximately doubled over
the interval $z\sim6$ to $z\sim3$, consistent with hierarchical
expectations.  Remarkably, this shift to lower luminosities extends to
even higher redshifts.  Using all deep $J+H$ NICMOS observations (800
orbits in total), we have been able to demonstrate that the bright end
of the LF ($>0.3L_{z=3}^{*}$) is at least 5 times lower at $z\sim10$
than at $z\sim4$, with a similar deficit being established from our
recent detections and first statistical sample of $z\sim7-8$ galaxies
using our UDF NICMOS data.  In these proceedings, we discuss what is
known about the UV LF and UV luminosity density at $z\sim6-10$ from current
data and its evolution relative to $z\sim3$.  We also describe several
exciting prospects for advance in this area over the next year.
\end{abstract}
 
\section{Introduction} 
Great progress has been made over the past few years in our
observational understanding of galaxies at the end of the first
billion years of the universe.  Much of this progress has come at
$z\sim3-6$ and has been due in large part to the substantial gains in
surveying efficiency made in the optical by the Advanced Camera for
Surveys (ACS) on HST.  This instrument provides us with a nearly
$\sim10\times$ increase in optical imaging efficiency over what was
available with WFPC2, higher resolution imaging, and a reasonably
efficient, red $z$-band filter.  Putting together these capabilities
with the well-established dropout technique (e.g., [27]), it became
possible to select literally hundreds to thousands of galaxies at
redshifts of $z\sim3-6$ \cite{bouw04b,giav04}.

\section{$z\sim6$ Rest-Frame Continuum UV LF}

Experience over the past two years has shown that a simple $i-z>1.3$
color selection is remarkably effective at isolating galaxies at
$z\sim6$ \cite{stan03,bouw03b,dick04}.  Though a small fraction of the
selected sources are found to be low mass stars and low-redshift
galaxies, most of these contaminants can be eliminated by making some
requirement on their stellarity (i.e., how pointlike they are) or
their flux in a bluer band (e.g., the $V_{606}$-band).  These results
have now been verified spectroscopically down to $z_{850,AB}\sim27.5$
\cite{malh05,stan04,dowh05}.

\begin{figure}
\begin{center}
\includegraphics[width=.54\textwidth]{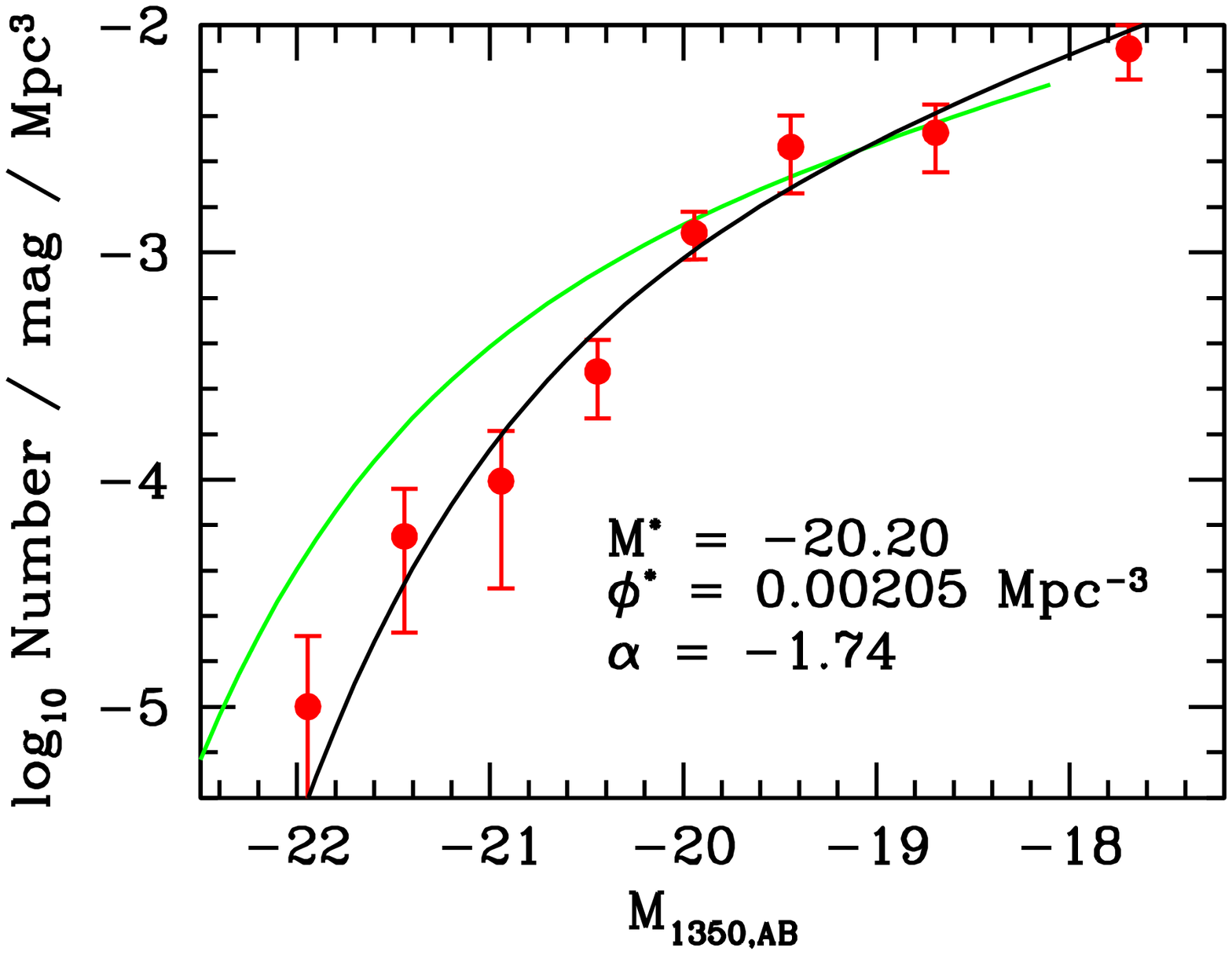}
\includegraphics[width=.45\textwidth]{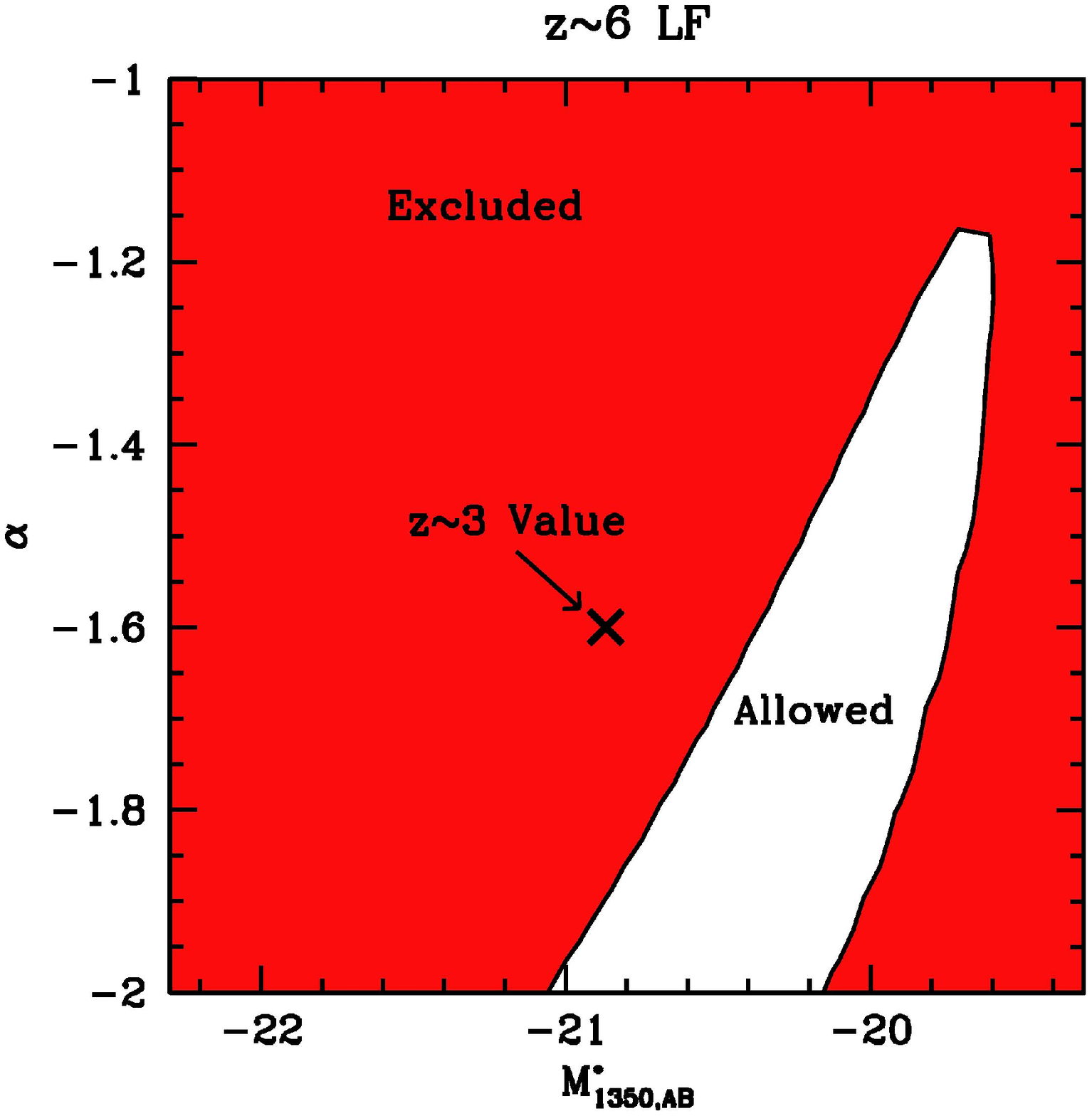}
\end{center}
\caption{\textit{(left panel).}  The rest-frame continuum $UV$
($\sim1350\AA$) LF estimated from the UDF, the UDF-Ps, and the GOODS
fields, shown in terms of the best-fit stepwise parameterizations
(\textit{red circles} with $1\sigma$ errors) and Schechter function
(\textit{solid black line})\cite{bouw05d}.  The $z\sim3$ LF is shown
for comparison (\textit{thick green line}) \cite{stei99}.  Our
$z\sim6$ LF shows a clear turnover at the bright end relative to the
$z\sim3$ LF and suggests that there has been a shift in the
characteristic luminosity from $z\sim6$ to $z\sim3$.  \textit{(right
panel).}  The allowed regions of parameter space (95\% confidence) for
the $z\sim6$ LF using current HST data \cite{bouw05d}.  The equivalent
$z\sim3$ values are shown with the large black cross.  Though only
modest constraints are possible on the $z\sim6$ faint-end slope
$\alpha$, the characteristic luminosity at $z\sim6$ appears to be
$\sim0.7$ mags fainter than at $z\sim3$.\label{fig:ilf}}
\end{figure}

With this increasing understanding of the $i$-dropout selection,
different studies attempted to construct rest-frame continuum $UV$
($\sim1350\AA$) luminosity functions (LFs) at $z\sim6$
\cite{bouw04a,bunk04,yan04,malh05}.  Despite small differences in
detail, each of these studies find that there were significantly fewer
galaxies at $z\sim6$ (per volume element) than at $z\sim3$.  A recent
determination of this LF by our team is shown in Figure~\ref{fig:ilf}a
and incorporates the results from some $\sim510$ $i$-dropouts selected
from the two GOODS fields (enhanced to include the SNe search data --
to be GOODS v2.0), two ACS parallel fields to the UDF, and the UDF
\cite{bouw05d}.  For comparison, we have also plotted the $z\sim3$ LF
\cite{stei99}, and it is amazing to note that the $z\sim6$ LF extends
fainter than at $z\sim3$, demonstrating the remarkable surveying
efficiency of current optical instrumentation.  It is also evident
that our $z\sim6$ LF shows a much larger deficit at the bright end
than at the faint end, suggesting that higher redshift galaxies are of
much lower luminosity (on average).  This explains (at least in part)
why early searches down to brighter limiting magnitudes \cite{stan03}
found more substantial evolution than similar searches down to fainter
magnitudes \cite{bouw03b,giav04}.

\begin{figure}
\begin{center}
\epsfig{figure=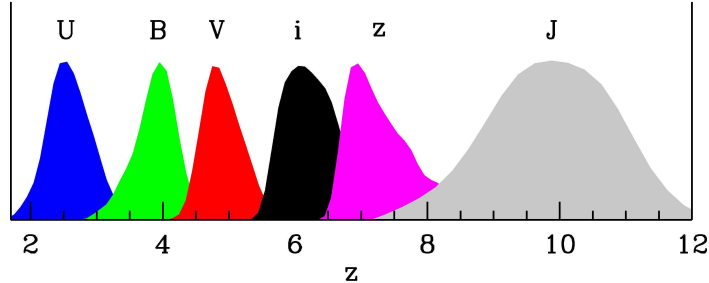,height=1.5in}
\end{center}
\caption{Estimated redshift distributions for fairly generic $U$, $B$,
$V$, $i$, $z$, and $J$-dropout selections.  The standard suite of HST
bands (e.g., F300W, F435W, F606W, F775W, F850LP, F110W) are used for
these selections.\label{fig:reddist}}
\end{figure}


We can look at this evolution more quantitatively.  In
Figure~\ref{fig:ilf}b, we plot the likelihood contours for the
$z\sim6$ LF and contrast it with the equivalent values at $z\sim3$
(thick black cross).  The red shaded regions on this plot indicate
those values of $\alpha$ and $M_{UV}^{*}$ which are ruled out at 95\%
confidence using the current HST data.  Though the constraints on the
faint-end slope $\alpha$ at $z\sim6$ are still somewhat modest, the
preferred values for the characteristic luminosity $M_{UV}^{*}$ are
much fainter than at $z\sim3$, suggesting that there has been
evolution in the characteristic scales on which star formation is
occurring.  This provides us with one of our first, most direct
evidences for the hierarchical buildup of galaxies early in the
history of the universe (see also [14] and [29]).  Further refinements
to the $z\sim6$ LF should be forthcoming over the next year due to the
availability of two additional deep fields to be taken with ACS
(GO-10632).


\section{First Detections of $z\sim7-8$ Galaxies}

Extending the dropout search beyond $z\sim6$ requires the detection of
objects in the infrared.  This has been difficult due to the
well-known limitations of infrared technology and because $z\sim7-8$
objects are likely to be very faint.  At $z\sim6$, the characteristic
luminosity is nearly $0.7$ mags fainter than at $z\sim3$.
Extrapolated to even higher redshift, these luminosities should be
even fainter still, perhaps 0.3 $L_{z=3}^{*}$.  In the $H$-band, this
corresponds to a magnitude of 27.3, which can only be reached in deep
NICMOS studies and very deep ground-based studies around massive
lensing clusters.

The deep NICMOS imaging over the optical UDF provided us with one of
our first opportunities to find $z\sim7-8$ galaxies.  The $5\sigma$
limiting depths in these data were 27.6 in the $J_{110}$-band and 27.4
in the $H_{160}$-band.  Moreover, the optical data for this field were
more than sufficient to set strong constraints on the $z_{850}$-band
fluxes.  Using a relatively aggressive set of detection criteria, we
carried out a $z_{850}$-dropout selection on these data and found 5
$z\sim7-8$ candidates.  Successive tests on our selection--including
scattering experiments and selection on the negative images--suggested
that most of our 5 candidates were likely at $z\sim7-8$ and there was
only one contaminant.  We therefore adopted as our likely sample 4
fiducial candidates.

To assess the implications of this first statistical sample of
$z\sim7-8$ objects, we generated no-evolution predictions based upon a
lower redshift $z\sim3.8$ $B_{435}$-dropout sample\cite{bouw05b} using
our well-established cloning machinery \cite{bouw98,bouw03a,bouw05c}.
An important consideration in projecting these lower redshift samples
to high redshifts was the observed evolution in size
\cite{ferg04,bouw04b} and UV color \cite{stan05,bouw05d}.  Running
through these simulations, we estimated that 14 objects would be found
(if there was no-evolution from $z\sim3.8$).  We compared this
prediction with our 4 fiducial $z$-dropout candidates, given the
expected small but non-zero contamination.  This suggested that the
rest-frame UV ($\sim1600\AA$) luminosity density at $z\sim7-8$ was
just $0.28\times$ that at $z\sim3.8$ (number weighted) or $0.20\times$
the $z\sim3.8$ value (if we use a luminosity weighting).

Though a first estimate of the rest-frame $UV$ luminosity density at
$z\sim7-8$, our determination still suffered from some substantial
uncertainties, notably the Poissonian errors ($\pm50$\%), cosmic
variance (factor of 2), as well as the overall contamination level
($\sim0-2$ objects).  This situation should improve substantially over
the next year using data from two HST programs: (1) a deep
$z_{850}$-dropout search in the field (GO-10632) and (2) a similar
search around seven massive lensing clusters (GO-10504 and GO-10699).
Both should yield $\sim5-10$ $z\sim7-8$ candidates, substantially
reducing uncertainties from our previously quoted estimates based on
the HUDF NICMOS footprint.  Simultaneously, searches with large
ground-based telescopes are ongoing and have yielded a sizeable number
of candidates, particularly around lensing clusters
\cite{mann05,rich05}.

\section{Searches for $z\sim10$ Galaxies}

The detection and confirmation of galaxies at $z\sim10$ appears to be
significantly more difficult than even at $z\sim7-8$.  Though the
additional distance plays a small role, by far the biggest challenge
is their luminosity: $z\sim10$ galaxies are expected to have very low
luminosities, several times lower than at $z\sim6$ or $z\sim7-8$
\cite{coor05,dave05}.  Even assuming no evolution in luminosity from
$z\sim7-8$ to $z\sim10$, typical magnitudes for these objects would be
$\sim$28 in the $H$-band, suggesting that one would need to probe to
very faint magnitudes indeed.

One way of reaching such magnitudes is to use the very deep IR imaging
capabilities of the HST NICMOS camera.  Unfortunately, even with this
instrument, searches for $z\sim10$ objects are still extremely
expensive, e.g., $\sim150$ orbits are needed to detect just one
$z\sim10$ object, and this assumes that the UV LF does not evolve from
$z\sim10$ to $z\sim6$.  It is thus no surprise that there has been a
lack of dedicated searches.  Nevertheless, it is possible to take
advantage of the deep $J_{110}$ and $H_{160}$ parallel data associated
with several HST deep fields to conduct a search.  Such data is
available for the HDF-S (150 orbits) and the UDF ($\sim$340 orbits).
There is also deep NICMOS J+H imaging over the WFPC2 HDF-N (176
orbits) and a portion of the ACS UDF (144 orbits).  The $5\sigma$
limiting magnitudes of these data range from $H_{160,AB}\sim27$ to
$H_{160,AB}\sim28.5$, which is sufficient to identify $z\sim10$
galaxies down to $0.3 L_{z=3}^{*}$.

To see what we could learn from these data, we carried out a search
for $(J_{110}-H_{160})_{AB}>1.8$ $J$-dropouts and found 11 objects
\cite{bouw05a}.  We eliminated those objects with $2\sigma$ detections
in the optical bands or with $H-K$ colors substantially redder than
the typical starburst type object (i.e., UV continuum slopes
$\beta>0.5$).  This left us with 3 candidates, all of which are in the
NICMOS parallel fields to the UDF.  Since we did not have comparably
deep images blueward or redward of the $J_{110}H_{160}$ passbands, it
was difficult for us to be very sure about the redshifts of our 3
candidates.  Nevertheless, we could still compare these 3 candidates
with the numbers expected assuming no-evolution from $z\sim6$.
Repeating the simulations described in \S3, we estimated that 4.8
$J_{110}$-dropouts would be found, only somewhat higher than our 3
candidates.  If we assume that all three candidates were at $z\sim10$,
the normalization we would obtain for the $z\sim10$ LF is just
$0.7\pm0.3\times$ that at $z\sim6$.  On the other hand, assuming that
none of these candidates were at these redshifts, the normalization we
would obtain is just $<0.2\times$ the $z\sim6$ value ($1\sigma$).
Remarkably, we found that these results were not very sensitive to
cosmic variance ($\sim19$\% RMS).  This is due to the extremely large
comoving distances probed in $J$-dropout searches ($\sim500$ Mpc).

\begin{figure}
\begin{center}
\epsfig{figure=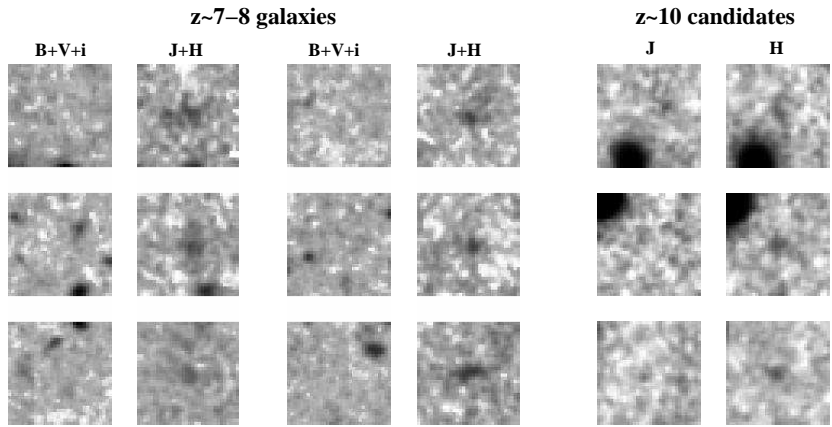,height=2.2in}
\end{center}
\caption{Candidate $z\sim7-8$ and $z\sim10$ galaxies from the UDF
NICMOS footprint and NICMOS parallels to the
UDF \cite{bouw04c,bouw05a}.\label{fig:highzcand}}
\end{figure}


The current round of HST proposals should strengthen these constraints
significantly.  Not only will we increase the total NICMOS observing
time relevant to these searches by $\sim60$\% (improving the
statistics), but we will obtain some very deep ACS observations
($\sim28.5$ mag at $5\sigma$ for apertures that match the NICMOS data)
over the deep NICMOS parallels containing our 3 $z\sim10$ candidates.
Independent searches are also ongoing around lensing clusters
(GO-10380, GO-10504, GTO-10699, and [22]), and we note that there are
claims by some teams to have a very good set of candidates.

\section{Summary}

Current observations are now of sufficient quality to robustly
determine the rest-frame $UV$ ($\sim1350$\AA) luminosity function at
$z\sim6$.  These determinations extend to nearly three magnitudes
below $L^{*}$, fainter than has been possible at $z\sim3$,
demonstrating the efficiency of current optical technology.
Substantial changes are evident in the LF relative to $z\sim3$,
suggesting that typical galaxies have more than doubled their star
formation rates over this interval.  The observed evolution is
suggestive of that expected from popular hierarchical models, and
would seem to indicate that we are literally witnessing the buildup of
galaxies over this range.

\begin{figure}
\begin{center}
\epsfig{figure=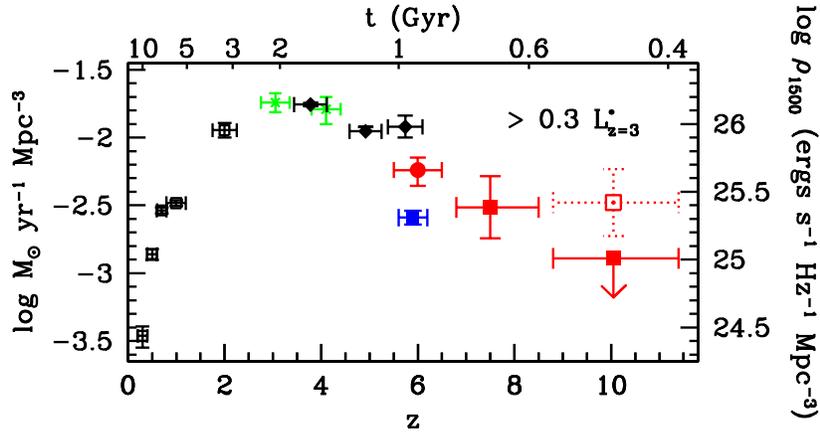,height=2.5in}
\end{center}
\caption{The cosmic star formation rate density versus redshift with
no extinction correction.  The star formation rate density (integrated
down to 0.3 $L_{z=3} ^{*}$ -- the limit of our $z\sim7-8$ and
$z\sim10$ searches) was calculated from the luminosity density in the
rest-frame UV continuum ($\sim1500\AA$: plotted on the right vertical
axis) using canonical assumptions and a Salpeter IMF \cite{mada98}.
The open red square at $z\sim10$ shows our result if 3 objects from
this study prove to be at $z\sim10$, while the large downward pointing
arrow shows our $1\sigma$ limits if none are \cite{bouw05a}.  Included
on this plot are also estimates by \cite{schi05} (open black squares),
\cite{stei99} (green crosses), \cite{giav04} (black diamonds),
\cite{bunk04} (solid blue square), \cite{bouw04c} (solid red square),
and \cite{bouw05d} (solid red circle).\label{fig:sfh}}
\end{figure}

Progress with higher redshift dropout samples has been less dramatic,
but is still forthcoming.  Our sample of $z\sim7-8$ $z_{850}$-dropouts
in the HUDF provided our first statistical sample of galaxies in this
epoch and thus allowed an initial estimate of the luminosity
density\cite{bouw04c}.  Searches for higher redshift $z\sim10$
$J$-dropouts have resulted in some candidates as well as yielding some
useful upper limits on the bright end of the $UV$ LF \cite{bouw05a}.
Galaxies at both times stand to undergo substantial improvements as
the result of future and ongoing programs.  Of course, for truly
substantial gains in this area, we will need to wait for the
availability of high resolution, high sensitivity space-based imagers,
such as WFC3 or NIRCam (JWST).

\acknowledgements{
We are indebted to the many members of the ACS GTO and UDF NICMOS GO
teams for their contribution to the current research.  Of particular
note was the assistance provided by John Blakeslee, Daniel Eisenstein,
Marijn Franx, Rodger Thompson, and Pieter van Dokkum, each of whom
contributed in an extremely important way.  We also acknowledge
valuable discussions with Brandon Allgood, Tom Broadhurst, Andy
Bunker, Akio Inoue, Sangeeta Malhotra, James Rhoads, Evan Scannapieco,
Daniel Schaerer, and Jason Tumlinson.  ACS was developed under NASA
contract NAS5-32865, and this research was supported under NASA grant
HST-GO09803.05-A and NAG5-7697.}

\def\Journal#1#2#3#4{{#1} {\bf #2}, #3 (#4)}
\def\mnras{{\em MNRAS}}

\vfill 
\end{document}